\documentclass [12pt]{article}

\usepackage{graphicx}

\small
      
\begin{document}
{\parskip 0cm

Contributed paper at the All-Russian Astronomical conference
"Close Binary Stars in Modern Astrophysics (MARTYNOV-2006)"
held in Moscow, Russia May 22 - 24, 2006

}
\begin{center}

\vspace{1.5cm}

{\Large\bf 
Possibilities of analysis of brightness distributions for
components of eclipsing variables from data of space
photometry }

\vspace{1cm}

{\it\bf M.B.Bogdanov$^1$, A.M.Cherepashchuk$^2$}

\vspace{1cm}

{\footnotesize 
$^1$Chernyshevskii University, Astrakhanskaya 83, Saratov, 410012 Russia

$^2$Sternberg Astronomical Institute, Universitetskii pr. 13, Moscow, 119992
Russia
}

\end{center}

\vspace{1cm}

We carried out numerical experiments on the evaluation of the
possibilities of obtaining the information about brightness
distributions for the components of eclipsing variables from
the data of high-precision photometry expected for planned
satellites COROT and Kepler. We examined a simple model of the
eclipsing binary with the spherical components on circular
orbits and the linear law of the limb darkening. The solutions
of light curves have been obtained as by fitting of the
nonlinear model, into the number of parameters of which
included the limb darkening coefficients, so also by the
solution of the ill-posed inverse problem of restoration of
brightness distributions across the disks of stars without
rigid model constraints on the form of these functions. The
obtained estimations show that if the observational accuracy
amounts to $10^{-4}$ then the limb darkening coefficients can be
found with the relative error approximately $0.01$ . The
brightness distributions across the disks of components can be
restored also nearly with the same accuracy.


\vspace{0.5cm}

Corresponding author. E-mail: BogdanovMB@info.sgu.ru

\vspace{1.5cm}

\centerline{\bf 1. Introduction}

\vspace{1cm}

A study of eclipsing variable stars is at present the basic
information source about sizes of stars and brightness
distributions across their disks. Information about brightness
distributions is especially important, since it allows to test
the models of the stellar atmospheres independent of spectral
analysis data. The analysis of the light curve of eclipsing
binary is the classical problem of astrophysics. The methods of
solution of this problem are detailed and connected with the
names of the noted researchers: H.N.Russell, D.Ya.Martynov,
V.P.Tsesevich, J.E.Merrill, Z.Kopal, {\it etc}.

At present there are
two basic approaches to the solution of the problem: the
fitting of the nonlinear model with known laws of the limb
darkening for components of the eclipsing variable and the
solution of the ill-posed inverse problem of the restoration of
brightness distributions across the disks of stars. However,
the precision of a ground-based photometry ( the relative error
of flux measurements, $\epsilon > 10^{-3}$ ) substantially limits the
accuracy of the obtained results.

The planned launches of the
special satellites COROT and Kepler make it possible to expect
a considerable increase in the precision of photometry of bright
objects ( up to $\epsilon = 10^{-5}$ ). Under these conditions new
possibilities for the solution of the classical problem of
investigating the eclipsing variable stars are opened [1].

The purpose of our paper is estimation of the possibilities of
the determination of geometric parameters and brightness
distributions for the components of eclipsing variables from
the data of high-precision photometry both the model fitting
method and the method of restoration of brightness distributions.

\vspace{1cm}

\centerline{\bf 2. The model light curve}

\vspace{1cm}

It is known that in the case of spherical components with a
linear law of the limb darkening the problem of calculating of
a light curve for an eclipsing variable has the exact solution,
whose error is determined by the precision of the numerical
estimation of one-dimensional definite integrals. Therefore for
further analysis we chose the simple model of an eclipsing
variable with the following parameters: the angle of orbital
inclination, $i = 89^o.0$; the radius of the first component in
units of the orbital radius, $r_1 = 0.30$; the luminosity of the
first component, $L_1 = 0.30$; the limb darkening coefficient of
the first component, $x_1 = 0.50$; the radius of the second
component, $r_2 = 0.20$; the luminosity of the second component,
$L_2 = 0.70$; the limb darkening coefficient of the second
component, $x_2 = 0.30$; The luminosities of the components are
connected by the equation: $L_1 + L_2 = 1$ . 

It is known that the light losses in minima $(1-l)_i$ are
described by the phase functions
$$
 (1-l)_i=\alpha^i(p,k)(1-l_A) , \eqno (1)
$$
where $(1-l_A)$ - the light loss at the moment of
the internal contact of the disks, $k=r_2/r_1$, $\Delta =r_1(1+kp)$,
and $\Delta $ - distance
between centers of the stellar disks in units of the orbital
radius. These phase functions, for one's part, can be
expressed via main phases for occultations $\alpha^\prime(p,k)$ and
transits $\alpha^{\prime\prime}(p,k)$ [2]. The last two main phases are
estimated numerically via 
calculations of one-dimensional definite integrals which
depend on parameters $p$ and $k$ .

We have calculated the
definite integrals of phase functions by application of the
Gauss - Kronrod algorithm using the subroutine DQAGE from
the SLATEC FORTRAN program library. The precision of
calculation of the light curve $l(i,L_1,r_1,x_1,L_2,r_2,x_2,\theta )$,
where $\theta$ - the light phase $(0\le \theta \le 1)$, was
adopted to be $\epsilon = 10^{-12}$. The calculated
light curve for our model of the eclipsing variable is shown
in figure 1. In the primary minimum occurs the total eclipse,
in the secondary - the annular eclipse.

\vspace{1cm}

\centerline{\bf 3. The model fitting}

\vspace{1cm}

One of basic approaches to the analysis of the light curves
of the eclipsing variables is the fitting of the nonlinear
model with known laws of the limb darkening for its
components. As the input data we took our model light curve,
perturbed by the influence of random noise with the various
values of $\epsilon$ . The dispersion of noise was assumed to be
constant in the magnitude scale. The Gaussian pseudo-random
numbers $\Delta l_i$ with zero mean and the standard deviation equal
to unity are used for the generation of the noise. Thus, the
samples of perturbed light curve can be written as
$l^o_i=l^c_i(1+\epsilon \Delta l_i)$ , where $l^c_i$ 
 -  the samples of the calculated model light curve.

In both primary and secondary minimum we considered $N = 100$
equidistant samples of the perturbed light curve. The search
for optimal values of the model parameters leads to the
solution of nonlinear minimization problems. In case of the
primary minimum
$$
  \sum_{j=1}^N [l^o(\theta_j)-l^c(i,r_1,r_2,L_2,x_2,\theta_j)]^2=min \eqno (2)
$$
and 
$$
  \sum_{j=1}^N [l^o(\theta_j)-l^c(i,r_1,r_2,L_1,x_1,\theta_j)]^2=min \eqno (3)
$$
for the secondary minimum. We have used for solutions of
these problems the DNLS1 subroutine also from the SLATEC
library which minimizes the sum of squares of nonlinear
functions by a modification of the Levenberg - Marquardt
algorithm [3].

Light curve nonlinearly depends on a number of
parameters. This non-linearity can leads to the presence of
the local minima of residual. For checking this possibility
we carried out a number of numerical experiments on the
solution of the problems of minimization (2) and (3) with
different initial values of the model parameters. In these
experiments the initial values were differed from the
precise values up to two times in the side of an increase or
decrease. In all cases the solution well converged to precise
and, thus, the local minima not were discovered.

We carried out the model fitting to perturbed light curves
for three values of $\epsilon = 10^{-5}, 10^{-4}$, and $10^{-3}$ .
In each case where examined $100$ curves with different realizations
of the random noise. The obtained average values of the model
parameters and their standard deviations are given in
tables 1 and 2. As can be seen from these tables, the
geometric parameters and the limb darkening coefficients are
evaluated very accurately from data of the high-precision
photometry. As a whole, a standard deviation in the
estimation of a model parameter linearly decreases with the
decrease of the relative error of the registration of light
curve. If the observational accuracy of the space
photometry amounts to $10^{-4}$ ( by one order higher than
precision of ground-based photometry ) then the limb
darkening coefficients can be found with the relative error
approximately $0.01$ .

\vspace{1.5cm}

\centerline{\bf 4. The restoration of brightness distributions}

\vspace{1cm}

An alternative approach to the analysis of the light curves of
the eclipsing variables - the restoration of brightness
distributions across the stellar disks without rigid model
constraints on the form of these functions was developed by
Cherepashchuk {\it et al.} [4]. Let $I(\xi)$\ \ $(0\le \xi \le r_1)$
and $I(\rho)$\ \ $(0\le \rho \le r_2)$ are the
brightness distributions across the disks of first and second
component respectively. It can be shown that the light loss in
the first and the second minimum are described by integral
equations 
$$
  1-l_1(\Delta) = \int_0^{r_1} K_1(\xi,\Delta,r_2)I(\xi)d\xi \eqno (4)
$$
$$
  1-l_2(\Delta) = \int_0^{r_2} K_2(\rho,\Delta,r_1)I(\rho)d\rho \eqno (5)
$$

Expressions for the kernels of these integral equations can be
found in paper Cherepashchuk {\it et al.} [4] and monograph Tsesevich
{\it et al.} [2].

The equations (4) and (5) are the integral equations of
Fredholm's first kind. The solution of this integral equation
is an ill-posed problem in Hadamard's sense and requires
utilization of {\it a priori} information about sought function. It is
possible to assume that for majority stars with thin
photospheres the brightness distributions are non-negative,
monotonically non-increasing, convex upward functions. It is
known that the sets of functions of these types are the compact
sets. The search for the solution of an ill-posed problem on the
compact set of functions gives the unique and stable result
[5]. This {\it a priori} information is qualitative and imposes no
rigid model constraints on the form of the brightness
distribution. Nevertheless, this guarantees that the obtained
brightness distributions will approach their exact values as
the errors in the registration of the observed light curve
approach zero with exception of the points of discontinuity
of the functions [5,6]. The use of a large amount of {\it a priori}
information about the possible form of the brightness
distribution in accordance with the physics of phenomenon
enables us to achieve a solution with a high degree of
stability against the effects of random noise. 

Thus, as the solution of our problem can be taken the
non-negative, monotonically non-increasing, convex upward
functions $I(\xi)$ and $I(\rho)$ that minimize the following functions of
functions - the norms in the $L_2$ function space:

$$
  \Phi_1[I(\xi),r_1,r_2,i]=\parallel \int_0^{r_1} K_1(\xi,\Delta,r_2)
  I(\xi)d\xi-[1-l_1(\Delta)] \parallel _{L_2}   \eqno (7)
$$                       
$$
  \Phi_2[I(\rho),r_1,r_2,i]=\parallel \int_0^{r_2} K_2(\rho,\Delta,r_1)
  I(\rho)d\rho-[1-l_2(\Delta)] \parallel _{L_2}   \eqno (8)
$$                       

The problem of restoration of the brightness distributions
depends also on three free parameters: $i, r_1$ and $r_2$ , whose
values can be found by the minimization of the summary
residual. 

We carried out the numerical experiment on the restoration
of brightness distributions from our perturbed model light
curve for the value of $\epsilon = 10^{-4}$ . The special estimations
showed that the choice of the number of points $M = 1001$ of
an uniform grid along a radius with the integration for
Simpson's formula makes it possible to ensure an relative
error in the calculation of integrals (4) and (5) below to
$10^{-6}$ . These grids where used later on for the minimization
of functions of functions. We minimized the
summary residual for both minima
$$
  \Phi_1[I(\xi),r_1,r_2,i]+\Phi_2[I(\rho),r_1,r_2,i]=min \eqno (9)
$$

and searched for the global minimum of the residual sum of
squares by variation of three geometrical parameters:
$i, r_1$, and $r_2$ . For the minimization of expressions (7) and
(8) on the compact set of non-negative, monotonically
non-increasing, convex upward functions for various values
of the geometrical parameters we used a modified version
of the PTISR code written in FORTRAN [5,6]. This code
minimizes residual by the method of the projection of the
conjugate gradients on the selected set of functions. To
reduce the effect of roundoff errors, we transformed all real
variables used in the PTISR and its auxiliary subroutines
into double precision variables with $16$ significant digits in
their floating-point mantissas. Zero initial approximations
for brightness distributions were used in all cases with the
minimization of expressions (7) and (8).

The numerical experiments shown that global minimum of
expression (9) can be found surely enough. In figure 2
presented the summary residual depending on the angle of
orbital inclination $i$ for optimal values of radii $r_1$ and $r_2$ .
The values of the geometrical parameters correspond to the
global minimum of residual are following: $i = 89^o.16 \pm
0^o.02, r_1 = 0.29990 \pm 0.00002, r_2 = 0.20005 \pm 0.00002$ .
Values of errors are formal and equal to steps of the grids
used in the carrying out variation of the geometrical
parameters.

The samples of the perturbed light curve are presented by
circles for the primary minimum in figure 3 and for the
secondary minimum - in figure 5. By solid lines in these
figures are shown the curves correspond to the restored
brightness distributions. The samples of the restored
brightness distributions are shown by circles in figure 4
and 6, where also are presented by solid lines the precise
distributions. Only each tenth sample is shown in order to
avoid of imposition. As can be seen from figures 4 and 6,
the accuracy of restoration of the brightness distribution
proves to be sufficiently high practically on entire disk of
star. Unfortunately, on the edges of disks, at the points of
the discontinuity of a function, the solutions are noticeably
differed from precise values. This is explained by the
absence of the convergence of solution of ill-posed problem at
the points of discontinuity of a function.

\vspace{1cm}

\centerline{\bf 5. Conclusion}

\vspace{1cm}

We carried out numerical experiments on the evaluation of the
possibilities of obtaining the information about brightness
distributions for the components of eclipsing variables
from the data of high-precision photometry expected for
planned satellites COROT and Kepler. We investigated of both
approaches to analysis of light curves: the fitting of the
nonlinear model, into the number of parameters of which
included the limb darkening coefficients, and the solution of
the ill-posed inverse problem of restoration of brightness
distributions across the disks of stars with using {\it a priori}
information about the form of these functions.

It is shown that in both cases the analysis of high-precision
data of a space photometry makes it possible to obtain the well
concordant results. The standard deviations in the estimations
of the model parameters linearly decrease with the decrease of
the relative error of the registration  of the light curve. If
the observational accuracy of the space photometry amounts to
$10^{-4}$ (by one order higher than precision of ground-based
photometry) then the limb darkening coefficients can be found
with the relative error approximately 0.01 . This accuracy will
make it possible to easily distinguish the linear law of limb
darkening of the nonlinear and to use for its estimation the
fitting of more complex models of brightness distributions.
The accuracy of restoration of the brightness distribution
without rigid model constraints on the form of this function
proves to be sufficiently high practically on entire disk of
star. Unfortunately, on the edges of disks, at the points of
the discontinuity of a function, the solutions are noticeably
differed from precise values. The geometric parameters of
eclipsing variable, found at the search for the global minimum
of residual in case of the restoration of brightness
distributions, also prove to be close to the precise values.

\vspace{1cm}

\centerline{\bf References}

\vspace{1cm}

{\parskip 0cm

1. C.Maceroni and I.Ribas, E-print, astro-ph/0511171 (2005).

2. V.P.Tsesevich (Editor), {\it Eclipsing variable stars} (Nauka,
Moscow, 1971) (in Russian).

3. J.E.Dennis and R.B.Schnabel, {\it Numerical methods for
unconstrained optimization and nonlinear equations} (Prentice - Hall,
Inc., Englewood Cliffs, New Jersey, 1983).

4. A.M.Cherepashchuk, A.V.Goncharskii and A.G.Yagola,
Sov. Astron. {\bf11}, 990 (1968).

5. A.N.Tikhonov, A.V.Goncharsky, V.V.Stepanov and A.G.Yagola,
{\it Regularizing algorithms and a priori information} (Nauka,
Moscow, 1983) (in Russian).

6. A.M.Cherepashchuk, A.V.Goncharsky and A.G.Yagola,
{\it Ill-posed problems in astrophysics} (Nauka, Moscow, 1985)
(in Russian).

}

\newpage

\vspace{3cm}

{\parskip 0cm

Table 1. The average values of the model parameters and their
standard deviations obtained by the model fitting from the
primary minimum for different values of relative error of the
light curve.

\vspace{3mm}
\hrule
\vspace{2mm}
Para-\hspace{0.5cm}Precise\hspace{2.1cm}$\epsilon=10^{-5}$\hspace{3.6cm}$\epsilon=10^{-4}$\hspace{3.2cm}$\epsilon=10^{-3}$    

meter\hspace{0.5cm}value
\vspace{2mm}
\hrule
\vspace{3mm}

   i\hspace{1.5cm}$89^o.0$\hspace{1.0cm}$88^o.9992\pm0^o.0034$\hspace{2.0cm}$89^o.001\pm0^o.035$\hspace{2.1cm}$89^o.09\pm0^o.41$

   $r_2$\hspace{1.3cm}0.20\hspace{0.9cm}$0.1999989\pm0.0000081$\hspace{1.3cm}$0.200002\pm0.000085$\hspace{1.35cm}$0.20006\pm0.00073$

   $L_2$\hspace{1.2cm}0.70\hspace{0.7cm}$0.70000000\pm0.00000072$\hspace{1.0cm}$0.7000002\pm0.0000074$\hspace{0.9cm}$0.700018\pm0.000078$

   $x_2$\hspace{1.25cm}0.30\hspace{1.3cm}$0.29998\pm0.00024$\hspace{2.2cm}$0.3001\pm0.0027$\hspace{2.1cm}$0.302\pm0.023$

   $r_1$\hspace{1.3cm}0.30\hspace{0.9cm}$0.3000007\pm0.0000028$\hspace{1.4cm}$0.300000\pm0.000030$\hspace{1.25cm}$0.30000\pm0.00028$

\vspace{3mm}

\hrule

}

\vspace{3cm}

{\parskip 0cm

Table 2. The average values of the model parameters and their
standard deviations obtained by the model fitting from the
secondary minimum for different values of relative error of the
light curve.

\vspace{3mm}
\hrule
\vspace{2mm}
Para-\hspace{0.5cm}Precise\hspace{2.4cm}$\epsilon=10^{-5}$\hspace{3.2cm}$\epsilon=10^{-4}$\hspace{2.8cm}$\epsilon=10^{-3}$    

meter\hspace{0.5cm}value
\vspace{2mm}
\hrule
\vspace{3mm}

   i\hspace{1.5cm}$89^o.0$\hspace{1.5cm}$88^o.994\pm0^o.018$\hspace{2.0cm}$88^o.99\pm0^o.15$\hspace{1.85cm}$88^o.69\pm0^o.99$

   $r_1$\hspace{1.3cm}0.30\hspace{1.4cm}$0.300001\pm0.000029$\hspace{1.25cm}$0.30000\pm0.00026$\hspace{1.35cm}$0.3002\pm0.0028$

   $L_1$\hspace{1.2cm}0.30\hspace{1.4cm}$0.300025\pm0.000085$\hspace{1.25cm}$0.30005\pm0.00074$\hspace{1.35cm}$0.3035\pm0.0069$

   $x_1$\hspace{1.25cm}0.50\hspace{1.6cm}$0.49989\pm0.00055$\hspace{1.65cm}$0.4995\pm0.0052$\hspace{1.75cm}$0.487\pm0.056$

   $r_2$\hspace{1.3cm}0.20\hspace{1.4cm}$0.199996\pm0.000017$\hspace{1.2cm}$0.20000\pm0.00016$\hspace{1.4cm}$0.1995\pm0.0015$

\vspace{3mm}

\hrule

}

\begin{figure}[p]
\centering
\includegraphics[width=17cm]{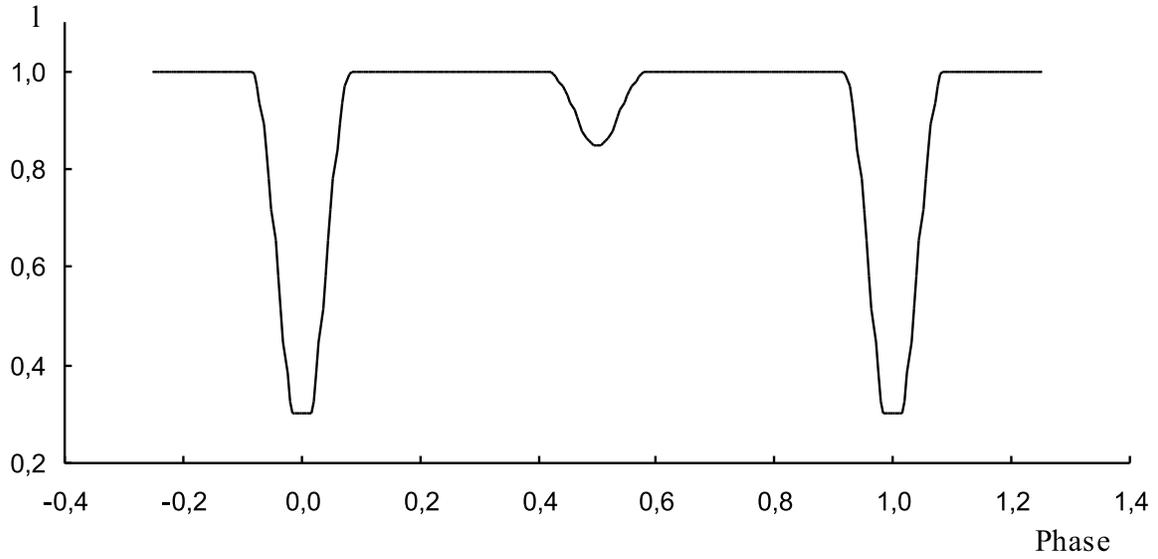}
\caption{The calculated light curve for chosen model of
the eclipsing variable.}
\label{fig1}
\end{figure}

\begin{figure}[p]
\centering
\includegraphics[width=17cm]{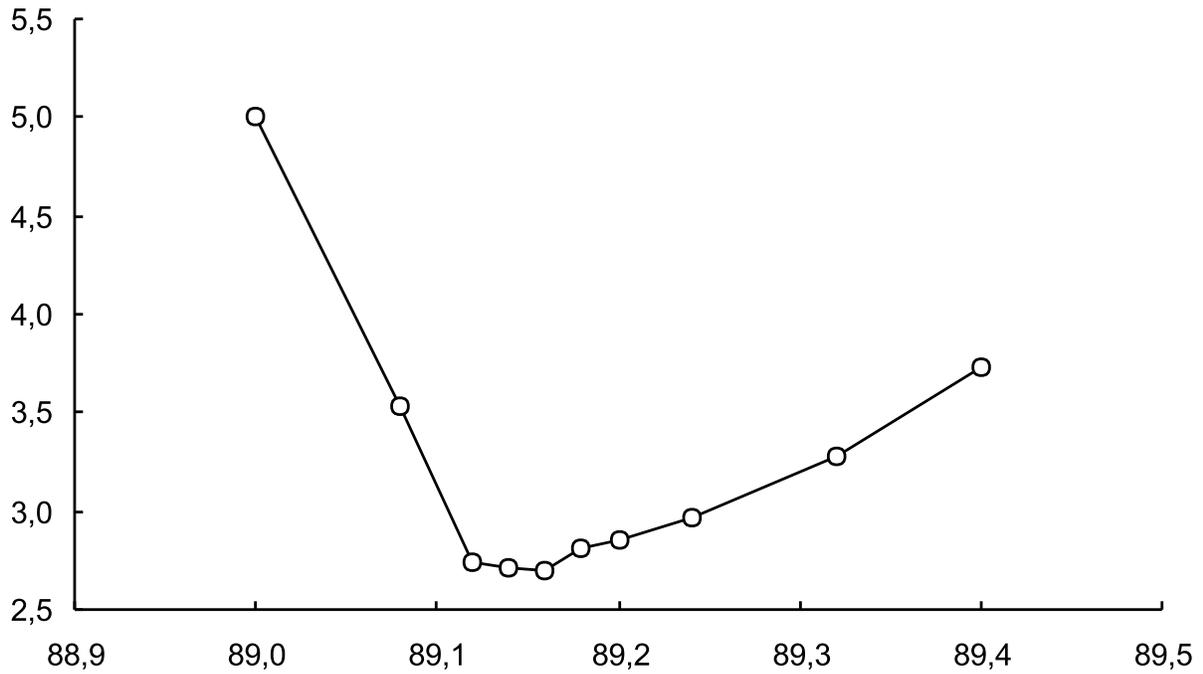}
\caption{The summary residual (in units $10^{-9}$) depending on
the angle of orbital inclination $i$ (in degrees) for optimal
values of radii $r_1$ and $r_2$ .}
\label{fig2}
\end{figure}

\begin{figure}[p]
\centering
\includegraphics[width=17cm]{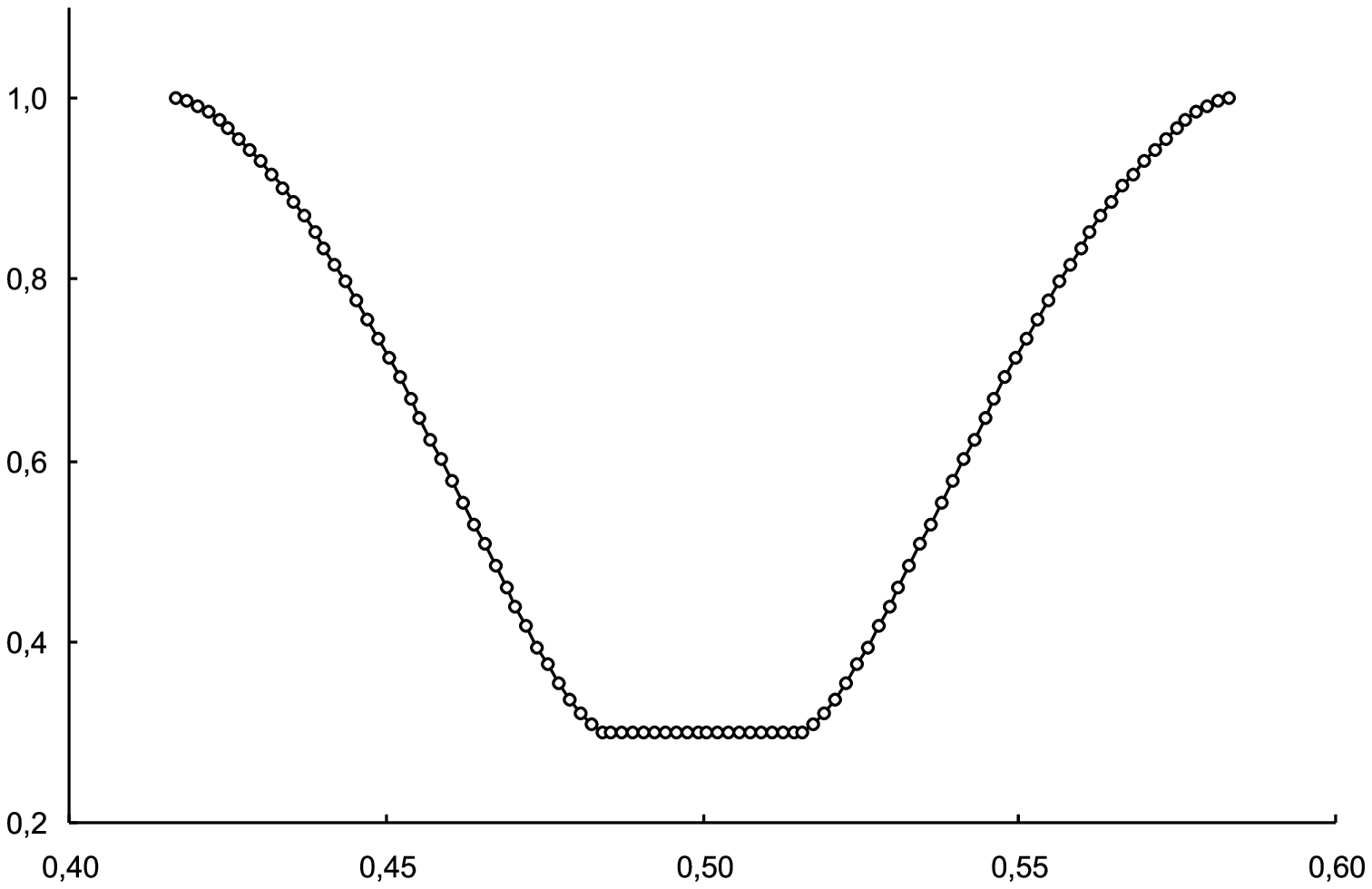}
\caption{The samples of the perturbed light curve for the
primary minimum (circles) and the light curve correspond to
the restored brightness distribution (solid line).}
\label{fig3}
\end{figure}

\begin{figure}[p]
\centering
\includegraphics[width=17cm]{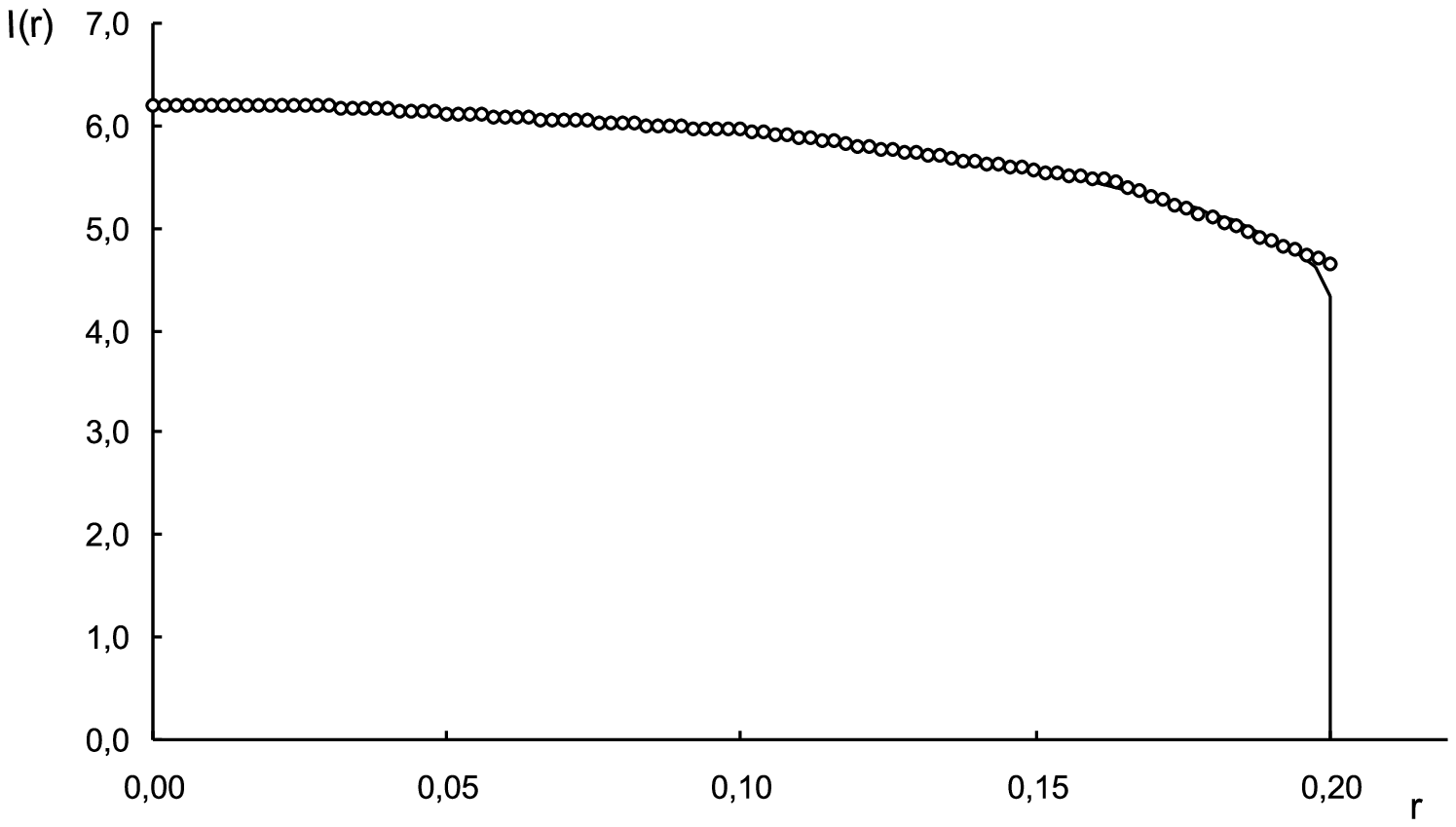}
\caption{The samples of the restored brightness
distribution for the second component (circles) and the
precise distribution (solid line).}
\label{fig4}
\end{figure}

\begin{figure}[p]
\centering
\includegraphics[width=17cm]{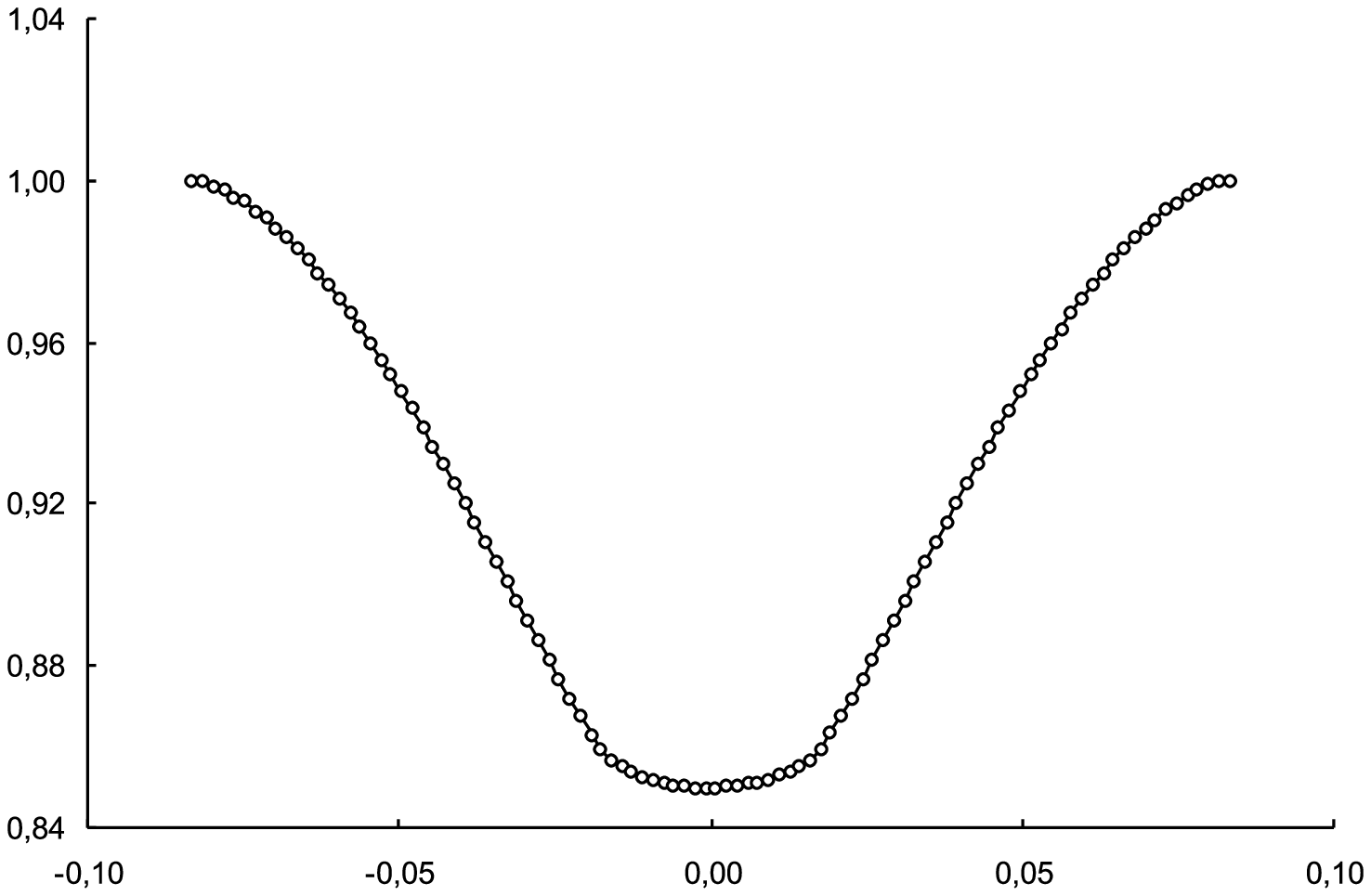}
\caption{The samples of the perturbed light curve for the
secondary  minimum (circles) and the light curve correspond
to the restored brightness distribution (solid line).}
\label{fig5}
\end{figure}

\begin{figure}[p]
\centering
\includegraphics[width=17cm]{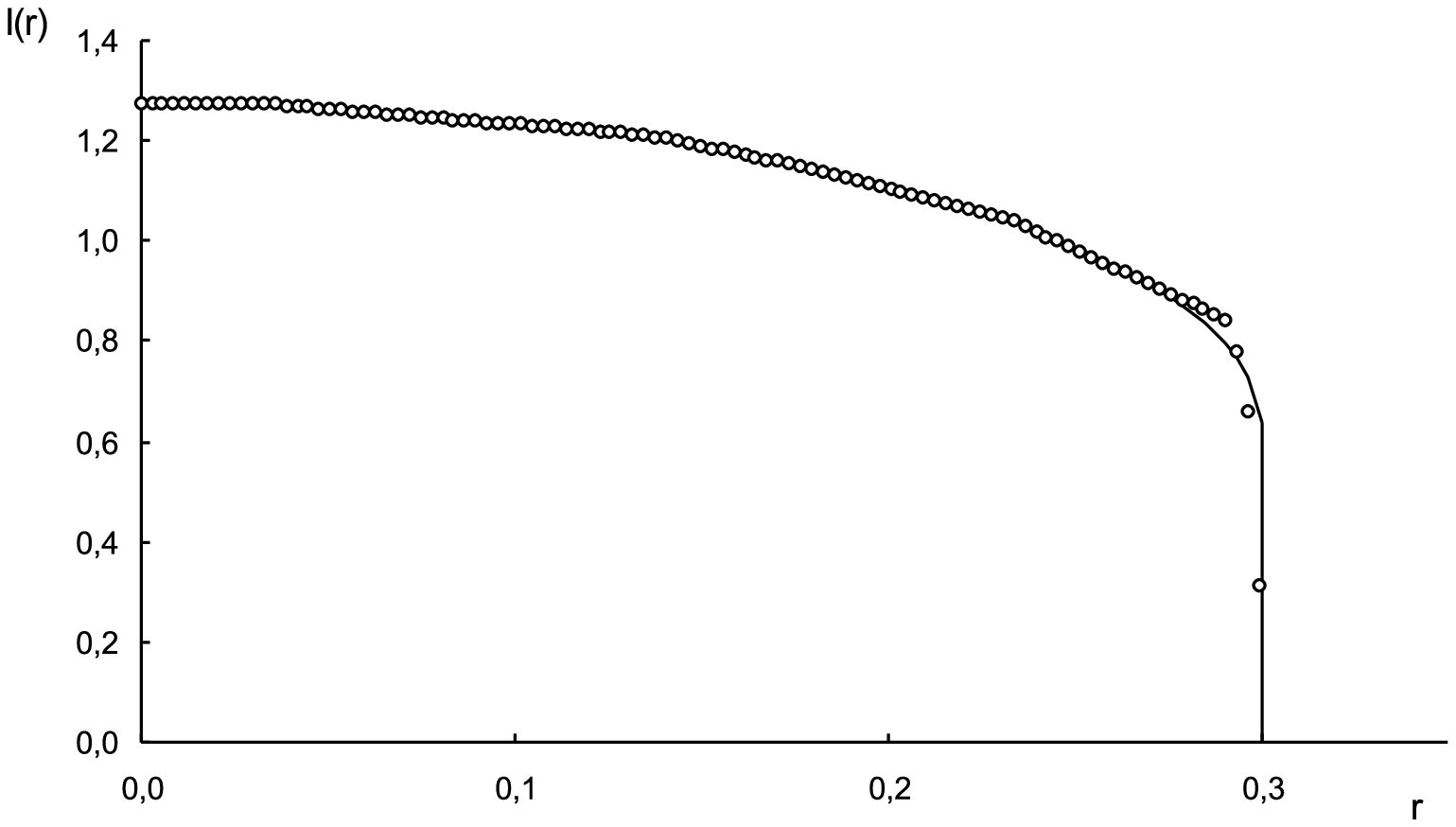}
\caption{The samples of the restored brightness
distribution for the first  component (circles) and the
precise distribution (solid line).}
\label{fig6}
\end{figure}

\end{document}